\documentclass[reprint,superscriptaddress,showpacs,amsmath,amssymb,prl]{revtex4-1}
\usepackage{mathtools}
\usepackage{natbib}
\usepackage{siunitx}
\usepackage{array}
\usepackage{xcolor}
\usepackage{subfigure} 
\usepackage{amsmath} 
\usepackage{amsfonts}
\usepackage{amssymb}
\usepackage{graphicx}
\usepackage{hyperref}

\providecommand{\vect}[1]{{\boldsymbol{#1}}}

\usepackage{physics}

\begin{document} 
\title{Voltage-Controlled High-Bandwidth Terahertz Oscillators Based On Antiferromagnets}

\author{Mike~A.~Lund}
\affiliation{Department of Engineering Sciences, University of Agder, 4879 Grimstad, Norway} 
\author{Davi R.~Rodrigues}
\affiliation{Department of Electrical and Information Engineering, Polytechnic University of Bari, 70125 Bari, Italy}
\author{Karin~Everschor-Sitte}
\affiliation{Faculty of Physics and Center for Nanointegration Duisburg-Essen (CENIDE), University of Duisburg-Essen, 47057 Duisburg, Germany}
\author{Kjetil M. D. Hals}
\affiliation{Department of Engineering Sciences, University of Agder, 4879 Grimstad, Norway} 
\date{\today}
\newcommand{\Kjetil}[1]{\textcolor{red}{#1}} 
\begin{abstract}
Producing compact voltage-controlled frequency generators and sensors operating in the terahertz (THz) regime represents a major technological challenge. Here, we show that noncollinear antiferromagnets (NCAFM) with kagome structure host gapless self-oscillations whose frequencies are tunable from 0 Hz to the THz regime via electrically induced spin-orbit torques (SOTs). The auto-oscillations' initiation,  bandwidth, and amplitude are investigated by deriving an effective theory, which captures the reactive and dissipative SOTs. We find that the dynamics strongly depends on the ground state's chirality, with one chirality having gapped excitations, whereas the opposite chirality provides gapless self-oscillations. Our results reveal that NCAFMs offer unique THz functional components, which could play a significant role in filling the THz technology gap.  
\end{abstract}

\maketitle 

The terahertz (THz) technology gap refers to a frequency range of electromagnetic radiation in the THz regime where current technologies are inefficient for generating and detecting radiation~\cite{Borak:sc2005,Tonouchi:np2007,Lee:sc2007}. While traditional electronics work well for producing and sensing microwaves and optics typically operate in the infrared region, few devices can utilize the THz range. THz devices are expected to have widespread applications ranging from improving the sensibility of biological and medical imaging techniques~\cite{Arnone:pw2000} to enhancing the functionality of information and communication technologies~\cite{Pang:jlt2022}. Therefore, developing compact and reliable THz components is one of the main challenges of today's electronics. 

In this context, antiferromagnetic spintronics has positioned itself as a promising future technology due to the intrinsic THz spin dynamics of antiferromagnets (AFMs)~\cite{Jungwirth:np2018,Duine:np2018,Gomonay:np2018,Zelezny:np2018,Nemec:np2018,Libor:np2018}. Notably, several works have demonstrated that the antiferromagnetic order couples to electric fields~\cite{Wadley:science2016,Reichlova:prb2015,Nunez:prb2006,Duine:prb2007,Gomonay:jmj2008,Wang:prl2008,Haney:prl2008,Gomonay:prb2010,Hals:prl2011,Gomonay:prb2012,Manchon:prb2014,Cheng:prb2014,Cheng:prl2014,Velkov:njp2016,Bodnar:prb2019,Cogulu:prb2021} – either indirectly via electrically generated spin currents or directly via spin-orbit torques (SOTs). This implies that it is possible to manipulate AFMs by electric fields and that AFMs can be used to modulate electric currents. Specifically, the latter effect has been proposed as a possible mechanism for developing nano-scale THz generators~\cite{Cheng:prl2016, Khymyn:scr2017, Zarzuela:prb2017, Troncoso:prb2019, Lisenkov:prb2019, Wolba:prb2021, Parthasarathy:prb2021,Zhao:prb2021,Ovcharov:pra2022}. 
The nano-oscillators use DC electric fields to create \emph{self-oscillations} in the AFM, which are sustainable cyclic modulations of the spin order driven without the stimulus of an external periodic force.
The self-oscillations act back on the electronic system, producing a THz electric output signal. Generally, there exists a frequency window in which both the amplitude and frequency of the AC output signal are tunable via the electric field. This frequency window represents the bandwidth of the nano-oscillators. The ability to maintain and control the self-oscillations over a broad range of frequencies is critical for the applicability of the nano-oscillators~\cite{Mohseni:sc2013, Akerman:nc2019}.

\begin{figure}[ht] 
\centering 
\includegraphics[scale=1.0]{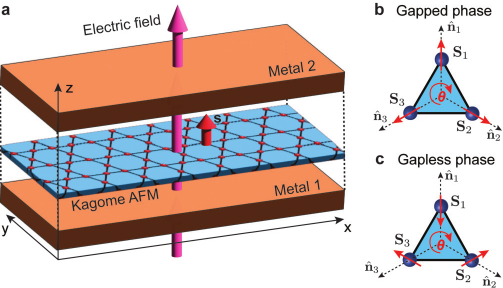}  
\caption{(color online). {\bf a.} A kagome AFM with broken mirror symmetry sandwiched between two metals.  
An electric field combined with spin-orbit coupling (SOC) generates an out-of-equilibrium spin density $\boldsymbol{s}$ collinear with the electric field, which can drive self-sustained oscillations in the AFM.  
{\bf b.}~({\bf c.}) A spin configuration with $(+)$-chirality  ($(-)$-chirality). The phase hosts gapped (gapless) self-oscillations corresponding to a rotation $\theta (t)$ of the sublattice spins about $z$. }
\label{Fig1} 
\end{figure} 

Previous works on AFM nano-oscillators have been theoretical and concentrated on so-called collinear AFMs~\cite{Cheng:prl2016, Khymyn:scr2017, Zarzuela:prb2017, Troncoso:prb2019, Lisenkov:prb2019, Wolba:prb2021, Parthasarathy:prb2021}, i.e., spin systems characterized by an antiparallel arrangement of the neighboring magnetic moments. 
However, in several AFMs, the spin sublattices are noncollinearly ordered. These spin systems are known as noncollinear AFMs (NCAFMs). In contrast to the collinear AFMs, where a staggered field parametrizes the spin order~\cite{Neel:AnnPhys1967}, a rotation matrix describes the spin order of NCAFMs~\cite{Andreev:spu1980}. Consequently, the NCAFMs exhibit more complex and intriguing spin physics than most ferromagnets and collinear AFMs. For example, recent works have revealed novel topological phenomena~\cite{Kuroda:nm2017} and a significant spin Hall effect~\cite{Kimata:science2004,Go:prl2022}. However, despite the great interest in NCAFMs, their current-driven self-oscillations remain largely unexplored~\cite{Zhao:prb2021}.

Here, we investigate the SOT-driven self-oscillations in a trilayer system consisting of a thin-film NCAFM with a kagome structure sandwiched between two metals. The external electric field is applied perpendicular to the thin-film plane (see Fig.~\ref{Fig1}a). Surprisingly, we find that the dynamics of the self-oscillations strongly depend on the chirality set by the relativistic Dzyaloshinskii–Moriya interaction (DMI) of the system. Despite the large in-plane and out-of-plane magnetic anisotropies, we show that one of the two chiral structures hosts gapless self-oscillations that are highly tunable via intrinsic SOTs. In contrast, the structure of opposite chirality has gapped oscillations. Notably, the gapless oscillations enable voltage-controlled NCAFM nano-oscillators with exceptional bandwidths, where the frequency is tunable from 0 Hz to the THz regime via the applied DC electric field. Our results thus demonstrate that the NCAFMs offer distinct chiral magnetic properties that are particularly attractive for
bridging the gap between technologies operating in the microwave and infrared regions. 

The material systems we consider are thin-film kagome AFMs, where the mirror symmetry of the kagome lattice plane is broken. 
These systems are described by the point group D$_6$~\cite{Hermele:prb2008}. 
Important candidate materials include Mn$_3$X (X= Ga, Ge, Sn), which in isolation are characterized by the point group ${\rm D}_{6h}$~\cite{Zhang:prb2017}, sandwiched between two different metals. 
The broken spatial inversion symmetry of the system has two significant consequences: 1) it leads to a magnetoelectric effect, and 2) it induces a DMI. The main effect of the DMI is that it determines the chirality of the ground state (see Fig~\ref{Fig1}b-c). The magnetoelectric effect refers to the out-of-equilibrium spin density produced by electric fields~\cite{Edelstein:ssc1990}, which in magnetic systems yields an SOT~\cite{Chernyshov:np2009,Hals:prb2013,Hals:prb2015, Zelezny:prb2017}. Below, we start by deriving the magnetoelectric effect of NCAFMs with D$_6$ symmetry from symmetry arguments~\cite{Comment:Symmetrygroup}. Then, based on the symmetry analysis, we phenomenologically add the coupling terms between the spin system and electric field in a microscopic model, which is used as starting point for deriving an effective action and dissipation functional of a uniform NCAFM. Further, the effective theory is applied to investigate the voltage-controlled self-oscillations. 

In linear response, the out-of-equilibrium spin density $\boldsymbol{s}$ produced by the electric field $\boldsymbol{\mathcal{E}}$ is given by~\cite{Edelstein:ssc1990}
\begin{equation}
s_i = \eta_{ij} \mathcal{E}_j. \label{Eq:Edelstein}
\end{equation}
Here, $\eta_{ij}$ is a second-rank axial tensor, which satisfies the following symmetry relationships~\cite{Hals:prb2013,Hals:prb2015}
\begin{equation}
\eta_{ij}= |\boldsymbol{G}| G_{ii^{'}} G_{jj^{'}} \eta_{i^{'}j^{'}} , \label{Eq:TensorSymmetry}
\end{equation}
dictated by the generators $\boldsymbol{G}$ of the system's point group.  
$|\boldsymbol{G}|$ represents the determinant of the symmetry operation $\boldsymbol{G}$. 
Throughout, we apply Einstein's summation convention for repeated indices. 
For kagome AFMs described by the point group $D_6$, the symmetry relations in Eq.~\eqref{Eq:TensorSymmetry} imply that $\eta_{ij}$ is diagonal and parameterized by two independent 
parameters~\cite{Birss:book}: $\eta_{xx}=\eta_{yy}\equiv \eta_{\bot}$ and $\eta_{zz}\equiv \eta_z$. 
Here, the $x$ and $y$ axes span the kagome plane, whereas the $z$-axis is perpendicular to the lattice plane (Fig.~\ref{Fig1}a).
Consequently, the out-of-equilibrium spin density produced by the electric field can be written as
\begin{equation}
\begin{pmatrix}
s_{x} \\
s_{y} \\
s_{z}
\end{pmatrix}
=
\begin{pmatrix}
\eta_{\bot} & 0 & 0 \\
0 & \eta_{\bot} & 0 \\
0 & 0  & \eta_{z}
\end{pmatrix}
\begin{pmatrix}
\mathcal{E}_{x} \\
\mathcal{E}_{y} \\
\mathcal{E}_{z}
\end{pmatrix} . \label{Eq:SpinDensity}
\end{equation} 
Interestingly, we see that the electric field in kagome AFMs can polarize the spin density along any axis (also the out-of-plane axis $z$).
This is different from most thin-film systems, which usually are characterized by Dresselhaus or Rashba SOC where the electric field only generates
spin densities polarized along an in-plane axis of the thin-film magnet~\cite{Chernyshov:np2009,Hals:prb2013,Hals:prb2015}. 
In what follows, we investigate how the spin density~\eqref{Eq:SpinDensity} couples to the NCAFM. 

The kagome AFM is modeled by the  spin Hamiltonian
\begin{equation}\label{Eq:Hamiltonian}
H = H_{e} + H_{a} + H_{D} + H_{\mathcal{E}}.
\end{equation}
Here, $H_{e}= J\sum_{\langle \iota \tilde{\iota} \rangle} \vect{S}_{\iota}\cdot\vect{S}_{\tilde{\iota}}$ describes the isotropic exchange interaction ($J > 0$) between the neighboring lattice sites $\langle \iota \tilde{\iota} \rangle$,  whereas 
 $H_{a}=  \sum_{\iota} [ K_z \left( \vect{S}_{\iota}\cdot\hat{\vect{z}} \right)^2 - K\left( \vect{S}_{\iota}\cdot\hat{\vect{n}}_{\iota} \right)^2]$ represents the easy axes ($K>0$) and easy plane ($K_z>0$) anisotropy energies. 
 The unit vector $\hat{\vect{n}}_\iota$ denotes the in-plane easy axis at lattice site $\iota$.
The kagome AFM consists of three spin sublattices with in-plane easy axes $\hat{\vect{n}}_1 = [0,1,0]$, $\hat{\vect{n}}_2 = [\sqrt{3}/2,-1/2,0]$, and $\hat{\vect{n}}_3 = [-\sqrt{3}/2,-1/2,0]$, respectively (Fig.~\ref{Fig1}b-c). 
$H_{D} = \sum_{\langle \iota \tilde{\iota} \rangle} \vect{D}_{\iota \tilde{\iota}}\cdot\left(\vect{S}_{\iota}\times\vect{S}_{\tilde{\iota}}\right)$ is the DMI where $\vect{D}_{\iota \tilde{\iota}}= D_z \hat{\vect{z}}$~\cite{Comment1}. 
$H_{\mathcal{E}}=  -\sum_{\iota}  g_r \vect{S}_\iota \cdot \boldsymbol{\eta} \boldsymbol{\mathcal{E}}$ expresses the reactive coupling to the electric field,
where $g_r$ is the coupling strength.

The ground state of the spin Hamiltonian~\eqref{Eq:Hamiltonian} depends on the ratio $D_z/K$.
If $D_z/K < 1/4\sqrt{3}$, the spins are aligned parallel or anti-parallel to the in-plane easy axes, i.e., $\vect{S}_{\iota}= \pm \hat{\vect{n}}_\iota$ (see Fig.~\ref{Fig1}b). 
We will refer to these two ground states as $(+)$-chiral.
On the other hand, if $D_z/K > 1/4\sqrt{3}$, the spins attain a configuration of opposite chirality, which we will refer to as having $(-)$-chirality (Fig.~\ref{Fig1}c).
The $(-)$-chiral configuration is related to $(+)$-chiral structure by a reflection about the $xz$-plane.     

The dynamics of the spin system is described by the action 
$\mathcal{S}= \sum_\iota \hbar \int {\rm dt} \vect{A} (\vect{S}_\iota) \cdot\dot{ \vect{S}}_\iota - \int {\rm dt} H $
and the dissipation functional
$\mathcal{G}= \sum_\iota \hbar \int {\rm dt} [ (\alpha_G/2) \dot{ \vect{S}}_\iota^2 + g_d \dot{ \vect{S}}_\iota \cdot (\boldsymbol{\eta}\boldsymbol{\mathcal{E}}\times \vect{S}_\iota ) ] $~\cite{Dombre:prb1989,Ulloa:prb2016,Rodrigues:prl2021,Lund:prb2021,Rodrigues:prb2022}.
Here, $\dot{ \vect{S}}_\iota\equiv \partial_t \vect{S}_\iota$, $\vect{A}$ is defined via $\vect\nabla \times \vect{A}(\vect{S}_\iota)   = \vect{S}_\iota/S$, $\alpha_G$ is the Gilbert damping parameter, and the term proportional to $g_d$ characterizes the dissipative coupling to the current-induced spin density.
To derive an effective description of the dynamics, it is convenient to express the three sublattice spins as~\cite{Dombre:prb1989}  
\begin{equation}\label{Eq:Representation}
\vect{S}_{\iota} (t) = \frac{S \mathbf{R} (t) \left[ \hat{\vect{n}}_{\iota} + a\vect{L} (t) \right] }{ \| \hat{\vect{n}}_{\iota} + a\vect{L} (t)  \| }, \ \ \iota \in \{ 1,2,3 \} .
\end{equation}
In Eq.~\eqref{Eq:Representation}, the rotation matrix $\mathbf{R}(t)\in SO(3)$ is the NCAFM's order parameter, whereas the vector $a\vect{L} (t)$ represents a spatial uniform small tilting (i.e., $\| a\vect{L} \| \ll 1$) of the spins.
The parameter $a$ is the lattice constant.

The effective action $\mathcal{S}_{\rm eff}$ for the order parameter $\mathbf{R}$ is obtained by
substituting Eq.~\eqref{Eq:Representation} into the action and expanding it to second order in the time variation $\dot{\mathbf{R}}$ and $a\vect{L}$~\cite{Dombre:prb1989,Ulloa:prb2016,Rodrigues:prl2021,Lund:prb2021,Rodrigues:prb2022,SupplMat}.
Minimizing the resulting action with respect to $\vect{L}$ yields an expression for the tilting field~\cite{SupplMat}  
\begin{equation}\label{Eq:L}
a \vect{L} = \boldsymbol{\Lambda} \mathbf{R}^T \left( \gamma_0 \boldsymbol{\omega} + \gamma_r \boldsymbol{\eta} \boldsymbol{\mathcal{E}}  \right) , 
\end{equation}
where $\gamma_0= \hbar/ 6 S J$, $\gamma_r= g_r/6SJ$, and $\boldsymbol{\Lambda}$ is a diagonal matrix with the elements $\Lambda_{xx}=\Lambda_{yy}= 2$ and $\Lambda_{zz}= 1$.
The vector $\boldsymbol{\omega}$ represents the angular velocity of the NCAFM and is governed by the time variation of $\mathbf{R}$:
\begin{equation}\label{Eq:omega}
\omega_i = -\frac{1}{2}\epsilon_{ijk} [\dot{\mathbf{R}} \mathbf{R}^T]_{jk} .
\end{equation}
The symbol $\epsilon_{ijk}$ denotes the Levi-Civita tensor.
Because $\vect{L}$ is fully determined by $\mathbf{R}$ and $\boldsymbol{\mathcal{E}}$, it is possible to eliminate the tilting field from $\mathcal{S}_{\rm eff}$ by substituting Eq.~\eqref{Eq:L} back into the action, which leads to the following expression in the continuum limit~\cite{SupplMat}  
\begin{equation}
\mathcal{S}_{\rm eff} =   \int {\rm dt dA} \left(   \frac{m}{2} \boldsymbol{\omega}^2  + \boldsymbol{\omega}\cdot \boldsymbol{\eta}_0\boldsymbol{\mathcal{E}} - \kappa_{klpn} R_{kl} R_{pn} \right). \label{Eq:Seff} 
\end{equation}
Here, $m= 2\hbar^2/\sqrt{3}J a^2$ is proportional to the moment of inertia of the AFM,  $\boldsymbol{\eta}_0= \hbar g_r \boldsymbol{\eta}/2 a_c J$, and the anisotropy tensor is $\kappa_{klpn}= \nu_{klpn} + d_{klpn}$ where
$\nu_{klpn}= \sum_{\iota=1,2,3} [ \tilde{K}_z n_{\iota l}n_{\iota n}\delta_{zp} \delta_{zk} -  \tilde{K} n_{\iota p}n_{\iota n} n_{\iota k}n_{\iota l} ]$ ($\delta_{ij}$ is the Kronecker delta)
and $d_{klpn}= (2/3\sqrt{3})\tilde{D}_z \epsilon_{zkp}[n_{1l} n_{3n} + n_{2l} n_{1n} + n_{3l} n_{2n} ]$.
The anisotropy constants are $\tilde{K}_z = K_zS^2/a_c$, $\tilde{K}= KS^2/a_c$, and $\tilde{D}_z= 3\sqrt{3}S^2 D_z/ a_c$ where $a_c= a^2\sqrt{3}/4$ is the area of the 2D unit cell. 
In Eq.~\eqref{Eq:Seff}, we integrate over the area of the thin-film AFM.

Using Eq.~\eqref{Eq:Representation}, a similar expansion of $\mathcal{G}$ to second order in $\dot{\mathbf{R}}$ and $a\vect{L}$ yields the effective dissipation functional~\cite{SupplMat} 
\begin{equation}
\mathcal{G}_{\rm eff} =  \int {\rm dt dA} \left(  \frac{\alpha}{2}  \boldsymbol{\omega}^2 + \beta\boldsymbol{\omega}\cdot\boldsymbol{\eta}_0 \boldsymbol{\mathcal{E}} \right), \label{Eq:Dissipation} 
\end{equation}
where $\alpha= 3\hbar S^2 \alpha_G/ a_c$ is the effective damping coefficient and the parameter $\beta= 6 S^2 J g_d/g_r$ expresses the ratio between the dissipative and reactive torques. 

Eqs.~\eqref{Eq:L}-\eqref{Eq:Dissipation} represent the first central result of this Letter and provide an effective theory of a kagome AFM coupled to an electric field via the intrinsic SOC.
The equations of motion follow from varying the action and dissipation with respect to $\mathbf{R}$.
In the following, we parameterize the rotation matrix by nautical angles~\cite{Haslwanter:book} 
\begin{equation}
\mathbf{R} =  \mathbf{R}_z (\theta)\mathbf{R}_y (\phi)\mathbf{R}_x (\psi) . \label{Eq:Rrepresentation} 
\end{equation}
Here, $\psi (t)$, $\phi (t)$ and $\theta (t)$ determine the rotation angles about the $x$, $y$, and $z$ axis, respectively.   
In this representation, the equations of motion of the AFM becomes
\begin{equation}
\frac{\delta\mathcal{S}_{\rm eff}}{\delta\vartheta} = \frac{\delta\mathcal{G}_{\rm eff}}{\delta\dot{\vartheta}}, \  \  \  \vartheta\in \{ \psi, \phi, \theta \} . \label{Eq:EOM0} 
\end{equation}

Next, we investigate how a DC electric field along $z$, i.e. $\boldsymbol{\mathcal{E}}= \mathcal{E}\hat{\vect{z}}$, can be applied to drive sustainable self-oscillations.
To this end, we first establish the electric threshold value $\mathcal{E}_{c}$ for initiating the self-oscillations before we determine how the electric field can be used to control 
the frequency and amplitude of the oscillations.

To derive $\mathcal{E}_{c}$, we consider small deviations away from the ground state and expand the action and dissipation to second order in the nautical angles. 
In this approximation, the anisotropy in Eq.~\eqref{Eq:Seff} can be written as $\kappa_{klpn} R_{kl} R_{pn}=  (1/2) \boldsymbol{\Omega}\cdot \mathbf{K}^{(\pm)}\boldsymbol{\Omega}$, whereas the
angular velocity in Eq.~\eqref{Eq:omega} becomes $\boldsymbol{\omega}= \dot{\boldsymbol{\Omega}} + (1/2)\boldsymbol{\Omega}\times\dot{\boldsymbol{\Omega}}$.
Here, $\boldsymbol{\Omega}= [\psi, \phi, \theta]$ and the tensor $\mathbf{K}^{(\pm)}$ is diagonal with the elements
$K_{xx}^{(+)}= K_{yy}^{(+)}= 3(\tilde{K} + \tilde{K}_z) - \tilde{D}_z$ and $K_{zz}^{(+)}= 6 \tilde{K}$
for the expansion around the state with $(+)$-chirality, and
$K_{xx}^{(-)}= K_{yy}^{(-)}= 3(\tilde{K} + 2\tilde{K}_z)/2 + \tilde{D}_z$ and $K_{zz}^{(-)}= 0$ for the state with $(-)$-chirality.
Varying the resulting action and dissipation functionals yields the linear equation:
\begin{equation}
m\ddot{\boldsymbol{\Omega}} = - \mathbf{K}^{(\pm)}\boldsymbol{\Omega} - \alpha\dot{\boldsymbol{\Omega}} + \dot{\boldsymbol{\Omega}}\times\boldsymbol{\eta}_0\boldsymbol{\mathcal{E}} - \beta[1- \frac{1}{2}\boldsymbol{\Omega}\times]\boldsymbol{\eta}_0\boldsymbol{\mathcal{E}} . \label{Eq:EOM} 
\end{equation}
 
We notice from Eq.~\eqref{Eq:EOM} that the ground state with $(-)$-chirality hosts gapless excitations because $K_{zz}^{(-)}= 0$.
The anisotropy constant $K_{zz}^{(-)}$ is zero because the Hamiltonian~\eqref{Eq:Hamiltonian} is invariant under rotation of the $(-)$-chiral state in the kagome plane~\cite{Comment:Gapless}.  
Thus, the excitations correspond to rotations of the spins by an angle $\theta$ about the $z$-axis (Fig.~\ref{Fig1}b-c). 
Importantly, the gapless excitations imply a zero threshold value $\mathcal{E}_{c}^{(-)}=0$ for initiating self-oscillations in the $(-)$-chiral state. 
This is surprising as the system is highly anisotropic with three in-plane easy axes as well as out-of-plane anisotropy.

In the $(+)$-chiral state, all excitations are gapped by the magnetic anisotropy.
To find $\mathcal{E}_{c}^{(+)}$, we substitute the ansatz $\boldsymbol{\Omega} (t) \sim \boldsymbol{\Omega}_0 \exp (i \omega t)$ into Eq.~\eqref{Eq:EOM} and solve the equation $\Im [\omega ( \mathcal{E}_{c}^{(+)} )]=0$~\cite{SupplMat}. 
Summarized, we find the following threshold values for the $(\pm)$-chiralities  
\begin{equation}
\mathcal{E}_{c}^{(\pm)}= \frac{2\alpha\sigma^{(\pm)} }{\eta_{0,zz}} \sqrt{ \frac{m K_{xx}^{(+)} }{ (m\beta-\alpha)^2 -\alpha^2} }, \label{Eq:EthP} 
\end{equation}
where  $\sigma^{(+)}= 1$ and $\sigma^{(-)}= 0$. 
For the $(+)$-chiral state, Eq.~\eqref{Eq:EthP} is supplied by the additional constraint $m\beta /\alpha \notin [0,2]$.  In the interval $m\beta /\alpha \in [0,2]$,  
the dissipative torque is incapable of destabilizing the ground state configuration and producing self-oscillations.  
Below the threshold value $\mathcal{E}_{c}^{(+)}$, the only effect of the electric field is to slightly rotate the ground state configuration by an angle
$\boldsymbol{\Omega}_{\ast} = -\beta \mathbf{A}^{-1}\boldsymbol{\eta}_0\boldsymbol{\mathcal{E}}$. 
Here, $A_{ij}= K^{(+)}_{ij} + \beta\epsilon_{ik j}\eta_{0,kk}\mathcal{E}_k/2$.

There also exists an upper critical value $\mathcal{E}_{f}^{(\pm)}$ where the electric field destroys the oscillations and drives the NCAFM  into a ferromagnetic phase~\cite{SupplMat,Comment2}:
\begin{equation}
\label{Eq:Ef} 
\mathcal{E}_{f}^{(\pm)} = \frac{S(6J + 2 K_z + K \mp 2\sqrt{3} D_z) }{ \eta_{zz} (g_r + \hbar g_d/\alpha_G)}.
\end{equation}

\begin{figure}[ht] 
\centering 
\includegraphics[scale=1.0]{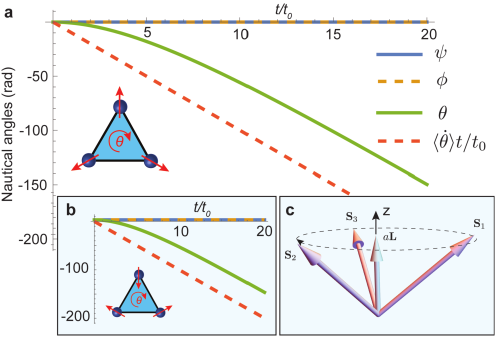}  
\caption{(color online). {\bf a.} ({\bf b.}) Auto-oscillation in a $(+)$-chiral ($(-)$-chiral) state. Obtained by solving the nonlinear equations of motion \eqref{Eq:EOM0} based on Eqs.~\eqref{Eq:Seff}-\eqref{Eq:Dissipation} with
the parameter values: 
$J=10$~meV, $K=0.03$~meV, $K_z=0.09$~meV, $D_z=K/8\sqrt{3}$ ($D_z=K/2\sqrt{3}$), $S=2.0$, $\alpha_G= 0.01$, $t_0= \hbar/S\sqrt{J(K+K_z)}$, $\beta t_0= 10$, $\mathcal{E}= 2 \mathcal{E}_c^{(+)}$.
The red line is the approximate solution $\theta (t)= \langle \dot{\theta}\rangle^{(\pm)} t$ based on Eq.~\eqref{Eq:dthetadt}.
{\bf c.} The self-oscillations are characterized by an electrically controllable out-of-plane tilting $a\vect{L}$ of the sublattice spins.   }
\label{Fig2} 
\end{figure}

Further, we examine how the electric field can control the frequency and amplitude of the self-oscillations.
A numerical solution of the full nonlinear equation of motion~\eqref{Eq:EOM0} based on Eqs.~\eqref{Eq:Seff}-\eqref{Eq:Dissipation}, shows that both ground states with $(\pm)$-chirality, respectively, evolve into a steady state oscillation $\theta (t)$ around the $z$-axis with $\psi=\phi=0$ when $\mathcal{E} > \mathcal{E}_{c}^{(\pm)}$ (Fig.~\ref{Fig2}a-b).    
Consequently, both chiralities are above the threshold characterized by the single nautical angle $\theta (t)$ and  
the dynamics of the auto-oscillations are captured by the ansatz $\theta= \theta (t)$ and $\psi= \phi=0$.
Upon substitution into Eqs.~\eqref{Eq:Seff}-\eqref{Eq:Dissipation}, the ansatz yields the following equation of motion~\eqref{Eq:EOM0} 
\begin{equation}
\label{Eq:EOMtheta} 
m \ddot{\theta} = -3\sigma^{(\pm)} \tilde{K} \sin (2\theta) - \alpha\dot{\theta} -\beta\eta_{0,zz}\mathcal{E} .
\end{equation}
Eq.~\eqref{Eq:EOMtheta} is identical to the equation of a point mass $m$ experiencing the periodic potential $-(3\tilde{K}/2) \cos (2\theta)$, friction $- \alpha\dot{\theta}$, and dissipative force $-\beta\eta_{0,zz}\mathcal{E}$. 
Because $\sigma^{(-)}=0$, the steady-state frequency $\dot{\theta}$ of the $(-)$-chiral state is easily extracted from Eq.~\eqref{Eq:EOMtheta} as the terminal velocity where the friction 
balances the dissipative force.  
This terminal velocity also corresponds to the time-averaged frequency of the $(+)$-chiral state, which can be calculated by averaging Eq.~\eqref{Eq:EOMtheta} over one cycle. 
Hence, for both chiralities, the relationship between the average frequency and the driving electric field becomes
\begin{equation}
\label{Eq:dthetadt} 
\langle \dot{\theta} \rangle^{(\pm)} = -\frac{\beta\eta_{0,zz}\mathcal{E}}{\alpha} .
\end{equation} 
Note that the self-oscillations of the $(+)$-chiral state can be maintained by a lower electric field strength $\mathcal{E}_{0}$ than the field $\mathcal{E}_{c}^{(+)}$ required for initiating the oscillations.
A similar phenomenon also appears in collinear AFMs~\cite{Khymyn:scr2017}. 
At the sub-threshold field $\mathcal{E}_{0}$, the work done by the dissipative force $-\beta\eta_{0,zz}\mathcal{E}$ equals the energy loss due to friction for the slowest possible oscillation (i.e., the oscillatory motion where $\dot{\theta}= 0$ at the energy maxima of the potential $-(3\tilde{K}/2) \cos (2\theta)$). This requirement leads to the sub-threshold field
$| \mathcal{E}_0 | = 2\alpha \sqrt{6\tilde{K} }/\pi\sqrt{m} | \beta\eta_{0,zz} | $.
Thus, we find the following bandwidths of the auto-oscillations:
\begin{equation}
\langle \dot{\theta} \rangle^{(\pm)}  \in - \frac{\beta\eta_{0,zz}}{  \alpha} [ \sigma^{(\pm)}\mathcal{E}_0  , \mathcal{E}_f^{(\pm)}]. \label{Eq:BandWidth} 
\end{equation}
In the frequency intervals \eqref{Eq:BandWidth}, the oscillation gradually changes from a full in-plane rotation of the spins into a conical motion 
where the base radius of the circular cone depends on $\mathcal{E}$ (Fig.~\ref{Fig2}c).
The tilting out of the $xy$-plane (and thus the amplitude of the oscillation) is determined by the vector $a \vect{L}$, which in linear response becomes:
 \begin{equation}
\label{Eq:TiltingSupCrit} 
\langle a \vect{L} \rangle =   \left( \frac{\alpha\gamma_r \eta_{zz} -  \gamma_0 \beta\eta_{0,zz} }{\alpha} \right) \mathcal{E} \hat{\vect{z}} .
\end{equation}  
 
Eqs.~\eqref{Eq:dthetadt}-\eqref{Eq:TiltingSupCrit} are the second central result of this Letter and provide a novel theory of electrically tunable nano-oscillators based on kagome AFMs.

The current-driven auto-oscillations, described by Eqs.~\eqref{Eq:dthetadt}-\eqref{Eq:TiltingSupCrit}, hold great potential for generating THz voltage signals. These oscillations stem from the anisotropic magnetoresistance (AMR) effect, which occurs when time variations in the spin system modulate the longitudinal resistance. In our study, we anticipate that the longitudinal resistance of the NCAFM is influenced by the nautical angle $\theta$ and the tilting vector $a \vect{L}$. This relationship can be expressed as 
$\mathcal{R}(t) = \mathcal{R}_0 + \delta \mathcal{R} (t)$, where $\mathcal{R}_0$ represents the constant component of the longitudinal resistance, and $\delta \mathcal{R} (t)= \delta \mathcal{R} [\theta (t), a \vect{L} (t)]$ varies with time via $\theta$ and $a \vect{L}$. The time-varying term, $\delta \mathcal{R}$, generates an AC voltage signal $U_{ac}$ given by 
\begin{equation}
U_{ac} (t) = \delta \mathcal{R} (t) I_{dc} , \label{Eq:Uac}
\end{equation}
where $I_{dc}$ represents the applied direct electric current.
Note that in ferromagnets and collinear AFMs, the AMR effect only depends on the relative angle between the current and the order parameter vector, thus, implying a vanishing AC output signal for 
precessional modes having a constant angle with respect to the applied current (such as the auto-oscillation mode for the $(-)$-chirality).
In collinear AFMs, theoretical works have shown that an AC signal can be achieved via interfacial spin-filtering~\cite{Cheng:prl2016}, in-plane anisotropy~\cite{Khymyn:scr2017}, and domain wall structures~\cite{Ovcharov:pra2022}.
However, NCAFMs have, in general, a much more complex spin structure than ferromagnets and collinear AFMs, parametrized by an SO(3)-valued order parameter field.
The AMR of NCAFMs is therefore anticipated to have a highly nontrivial and anisotropic dependence on the orientation of the underlying spin-lattice. 
This has recently been experimentally demonstrated for Mn$_3$Ge~\cite{Qin:acs2020}. 
Consequently, we expect that even highly symmetrical auto-oscillation modes (such as the $(-)$-chirality mode) could potentially lead to modulations of the longitudinal resistance in Eq.~\eqref{Eq:Uac}. 

The frequency and bandwidth of the generated voltage signal~\eqref{Eq:Uac} are determined by the angular velocity~\eqref{Eq:dthetadt} and the frequency window~\eqref{Eq:BandWidth}, respectively. To estimate the characteristic frequency range of our nano-oscillator, we assume that the NCAFM’s SOT is comparable to that of (Ga,Mn)As~\cite{Kurebayashi:nn2014}, which yields the following values for the reactive and dissipative torque parameters: $g_r\eta_{zz}/\hbar= 879.3$~m/Vs and $g_d\eta_{zz}= 367.2$~m/Vs. By utilizing these values, along with the material parameters provided in Fig.~\ref{Fig2}, we find the bandwidths to be $\langle \dot{\theta} \rangle^{(-)}  \in [0  , 1.8\times 10^{14}]$~rad/s and $\langle \dot{\theta} \rangle^{(+)}  \in [3.7\times 10^{12}  , 1.8\times 10^{14}]$~rad/s, respectively, and an initiation frequency of $\sim 1.6\times 10^{13}$~rad/s for the $(+)$-mode~\cite{SupplMat}. 
These estimations demonstrate that NCAFM-based nano-oscillators offer a unique frequency tunability, which is challenging to achieve in other magnetic systems. It is noteworthy that NCAFMs exhibiting $D_4$ and $D_3$ symmetry possess the same spin density~\eqref{Eq:SpinDensity} and SOT as kagome AFMs. Consequently, we anticipate that these material classes will show similar current-driven auto-oscillations. 

KMDH acknowledges funding from the Research Council of Norway via Grant No. 286889.
KES acknowledges funding from the German Research Foundation (DFG) Project No. 320163632 and the TRR 173 – 268565370 Spin + X (project B12).
DRR acknowledges funding from the Ministero dell'Università e della Ricerca, D.M. 10/08/2021 n. 1062 (PON Ricerca e Innovazione) and Project PRIN: “The Italian Factory of Micromagnetic Modeling and Spintronics” (Prot. 2020LWPKH7).



\begin{thebibliography}{99}
\bibitem{Comment1}Note that the $D_6$ symmetry also allows for an in-plane component of the DMI vector, which produces a weak ferromagnetic phase.
In the present work, we consider an antiferromagnetic phase and thus disregard the in-plane component of the DMI.
\bibitem{Comment:Gapless} The spin configuration $\vect{S}_1^{(-)} = S [0,-1,0]$, $\vect{S}_2^{(-)} = S [\sqrt{3}/2,1/2,0]$, and $\vect{S}_3^{(-)} = S [-\sqrt{3}/2,1/2,0]$ describes the ground state with $(-)$-chirality.
Substituting $\vect{S}_{\iota} (\theta)= \mathbf{R}_z(\theta)\vect{S}_{\iota}^{(-)}$ into $H_a$ in Eq.~\eqref{Eq:Hamiltonian} yields $-3 S^2 K / 2 $.
This implies that the spin Hamiltonian~\eqref{Eq:Hamiltonian} is invariant under rotation of the $(-)$-chiral state in the kagome plane. 
Higher order terms of the form $\sim (\vect{S}_{\iota} \cdot \hat{\vect{n}}_{\iota})^n$ with $n=4,6,...$ are expected to stabilize the state with $(-)$-chirality with respect to thermal fluctuations. 
However, these terms are anticipated to be considerably smaller than the second order harmonics $\sim (\vect{S}_{\iota} \cdot \hat{\vect{n}}_{\iota})^2$.
\end{thebibliography}
\end{document}